# Mining Twitter Conversations around E-commerce Promotional Events

**Binny Mathew**
Dept. of CSE
IIT Kharagpur, India
binny.iitkgp@gmail.com

**Niloy Ganguly**
Dept. of CSE
IIT Kharagpur, India
niloy@cse.iitkgp.ernet.in

**Unnikrishnan T A**
Dept. of CSE
IIT Kharagpur, India
unnikrishnan.t.a@gmail.com

**Samik Datta**
Flipkart Ltd.
Bangalore, India
samik.datta@flipkart.com

**Tanmoy Chakraborty**
Dept. of CSE
IIT Kharagpur, India
its_tanmoy@cse.iitkgp.ernet.in

## Abstract
With Social Media platforms establishing themselves as the de facto destinations for their customers' views and opinions, brands around the World are investing heavily on invigorating their customer connects by utilizing such platforms to their fullest. In this paper, we develop a novel technique for mining conversations in Twitter by weaving together *all* conversations around an event into one unified graph (*Conversation Graph*, henceforth). The structure of the Conversation Graph emerges as a variant of the BOWTIE structure (dubbed ASKEWBOWTIE henceforth) as a result of the complex communication patterns amongst these players. Finally, we investigate the structural properties of the ASKEWBOWTIE structure to understand the configuration of the components and their temporal evolution.

## ACM Classification Keywords
E.1 [Data Structures]: Graphs and Networks

## Introduction
The open nature of Twitter and its wealth of publicly available communications have attracted the attention of the research community. However, majority of the existing literature [3] focus on studying these conversations *in isolation*. These works highlight the importance of studying the *overall* structure that emerges in Social Media due to an Event. Chierichetti et al. [2] studied short-lived events, like a goal scored in SUPERBOWL, where users share their views and opinions to the World. In contrast, in this paper we collect and study *all* the conversations that took place around three large-scale e-commerce promotional events conducted by India's largest e-commerce portal, Flipkart. Through a detailed study of the users' profile, activity etc., we find several classes of users existing in the conversational ecosystem. We weave together all these conversations into a *Conversation Graph*, and find that the graph demonstrates a stable BOWTIE-like structure - hitherto observed in World Wide Web [1] - mainly comprising of three connected components - IN, LSCC[1], OUT (Figure 1). The IN component enjoys the largest concentration of users, and hence, we term this variant the ASKEWBOWTIE. Finally, we conduct a thorough analysis to understand the structural properties of the ASKEWBOWTIE.

| Event | #tweets | #users | # tweets with @ |
|---|---|---|---|
| BBD | 135,593 | 40,891 | 96,263 |
| BASD2 | 51,323 | 10,926 | 43,360 |
| BASD3 | 160,322 | 19,155 | 139,896 |

**Table 1:** Description of the data-sets.

| Data-set | Precision% | Recall% |
|---|---|---|
| BBD | 89.50% | 89.82% |
| BASD2 | 90.54% | 91.56% |
| BASD3 | 92.36% | 92.54% |

**Table 2:** Accuracy of sentiment tagger for three datasets.

| Graph | $|\mathcal{U}|$ | $|\mathcal{T}|$ | |LSCC| | $|\mathcal{T}(\text{LSCC})|$ |
|---|---|---|---|---|
| BBD | 36K | 118K | 83% | 95.21% |
| BASD2 | 8K | 56K | 88% | 96.54% |
| BASD3 | 14K | 210K | 90% | 95.49% |

**Table 3:** Statistics of the Conversation Graphs. The columns |LSCC| and $|\mathcal{T}(\text{LSCC})|$ denote the percentage concentration of the total SCC mass and flow associated with LSCC respectively.

| Component | BBD | BASD2 | BASD3 |
|---|---|---|---|
| IN | 63% | 55.3% | 57.1% |
| LSCC | 12% | 23.1% | 21.1% |
| OUT | 1.5% | 3.5% | 1.2% |
| TENDRILS | 21% | 13.8% | 14.3% |
| DISC | 2.8% | 6.7% | 6.2% |

**Table 4:** Concentration of masses in different components of the ASKEWBOWTIE.

**Data-sets**

We focus on three large-scale promotional events, conducted by FLIPKART, that attracted enormous public and mainstream media attention: BigBillionDay (BBD), BigAppShoppingDays (BASD2 and BASD3). The statistics of the crawled data-sets are presented in Table 1. We further gathered from Twitter other publicly available information on the authors of those tweets, e.g., account creation time, total number of tweets till date, list of followers and followings, the time-stamp of their latest tweets etc. As an estimate of their *influence*, we also obtained their $Klout$-scores[2] through the publicly available APIs.

---
[1] Largest Strongly Connected Component
[2] https://klout.com/home

**Users and Opinions**

Given that $11\%$ of the tweets across our data-sets were written in vernacular[3], we chose not to use off-the-shelf Sentiment Analysis tools. Instead, building upon the features proposed in [5], we train a binary classifier (Support Vector Machine). Three annotators annotated $1000$ tweets for each data-set with positive and negative sentiments (average inter-annotator agreement $89\%$). On average, we obtain precision and recall of $90\%$, as can be seen from Table 2 with $10$-fold cross-validation.

Based on this Sentiment Analysis, we further classify the Twitter handles into $4$ categories: (i) HAPPY: Happy customers having a positive opinion on average, (ii) UNHAPPY: Unhappy customers who had negative opinion on average, (iii) ADVERSARIAL: Adversarial accounts spreading negative sentiments, *purposefully*, (iv) PROMOTER: Friendly accounts spreading positive sentiments, perhaps as a result of incentives. We resort to a collection of heuristics to arrive at a rule-based classifier for this purpose.

**The Conversation Graph**

Twitter provides two peer-to-peer communication primitives - reply and mention - that encourage conversations amongst the users. We also consider re-tweets, in our definition of conversation.

Let $\mathcal{T}$ and $\mathcal{U}$ denote the universe of such tweets and users, respectively, in our crawls. For each tweet, $t \in \mathcal{T}$, let $a(t) \in \mathcal{U}$ and $R(t) \subset \mathcal{U}$ denote the author and recipients of the tweet $t$, respectively. Then we create a *Conversation Graph*, $G(\mathcal{U}, \mathcal{T})$. For each tweet $t \in \mathcal{T}$, a directed edge is added from $u$ to $v$, whenever $u = a(t)$ and $v \in R(t)$. Multiple edges between vertex-pairs are converted to weighted edges. Furthermore, let $t_a$ denote the creation-time of the tweet. The sub-graph induced by tweets created during $(d, d + \Delta)$, $G(\mathcal{T}_{(d,d+\Delta)})$ allows us to view a time-slice of the conversation graph. For subsets of users, $U \subset \mathcal{U}$, we let $G(U)$ denote the sub-graph induced by vertices in $U$. Similarly, for a pair of such subsets, $(U_i, U_j)$, we let $\mathcal{T}(U_i \to U_j) = \{t \mid a(t) \in U_i, R(t) \cap U_j \neq \emptyset, t \in \mathcal{T}\}$ denote the *flow* across the *cut* $(U_i, U_j)$. We summarize all the conversation graphs in Table 3.

---
[3] https://github.com/irshadbhat/litcm

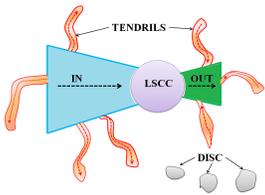

**Figure 1:** A Schematic diagram of a ASKEWBOWTIE.

| Data-sets | Measures | $\mathcal{T}(\text{IN} \to \text{LSCC})$ | $\mathcal{T}(\text{LSCC} \to \text{OUT})$ |
|---|---|---|---|
| BBD | WF | 27930.01 | 423.31 |
| | LD | 0.47 | 0.48 |
| | PP | 0.34 | 0.41 |
| | CW | 0.01 | 0.004 |
| BASD2 | WF | 9152.05 | 408.71 |
| | LD | 0.46 | 0.47 |
| | PP | 0.10 | 0.15 |
| | CW | 0.006 | 0 |
| BASD3 | WF | 24264.46 | 3463.02 |
| | LD | 0.46 | 0.49 |
| | PP | 0.20 | 0.70 |
| | CW | 0.02 | 0.002 |

**Table 5:** Formality of tweets in $\mathcal{T}(\text{IN} \to \text{LSCC})$ and $\mathcal{T}(\text{LSCC} \to \text{OUT})$. Features [4]: (i) Word Frequency (*WF*) measuring rarity of words, (ii) Lexical Density (*LD*) capturing the stylistic difference between corpora in terms of the proportion of verbs, nouns, adjectives and adverbs, (iii) Personal Pronouns (*PP*) measuring the usage of Third Person Pronouns, and, (iv) Curse Words (*CW*) measuring frequency of abusing words. Higher values of LD and PP hints at more formal communication; whereas WF and CW has opposite effects.

**The ASKEWBOWTIE.** To begin with, we decompose the graph into its Strongly Connected Components (SCC). All pairs of vertices within each SCC enjoy bi-directional connectivity. We observe that the largest such SCC (dubbed LSCC) is a giant, accounting for $81.5\%$ of SCC mass throughout our data-sets. We further decompose $G(\mathcal{U}, \mathcal{T})$ into components according to their reachability to and from LSCC. For example, all vertices in LSCC are reachable from all vertices in IN, but *not* vice-versa. Similarly, all vertices in OUT are reachable from all vertices in LSCC in a strictly uni-directional manner. Note that vertices in TENDRILS enjoy no directed reachability to and from LSCC. Vertices in DISC are disconnected from rest of the graph. A schematic of the structure is presented in Figure 1. The relative sizes of these components across our data-sets are captured in Table 4. Given the skew of concentration towards the IN component, we call the structure ASKEWBOWTIE in what follows.

Next, we study the users that fall into each of these components.

*Components*
To understand the components better, we now set out to investigate the users that fall into different components of ASKEWBOWTIE.

**User Type.** First, we study the distributions of different handle-types across components. We observe that, across our data-sets, majority of the PROMOTER handles fall into LSCC, whereas, majority of the ADVERSARIAL handles are concentrated within IN. Given the heavy concentration of users in IN, we observe a large concentration of handles of all types, apart from PROMOTER.

**User Influence.** To study the influence of the users, we use *Klout* scores as an estimate. The median influence score for BBD in LSCC and OUT is $66\%$ and $71\%$ higher than that of the other components, respectively. This conforms to our intuition than LSCC and OUT house more influential users than rest of the graph.

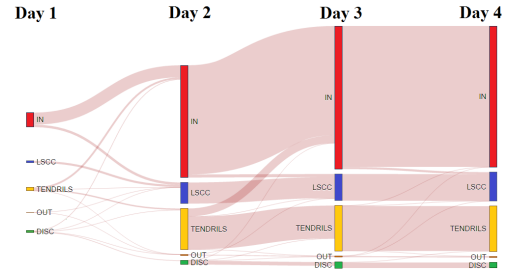

**Figure 2:** Alluvial diagram depicting the migration of users across components of the ASKEWBOWTIE over four consecutive days during BBD. Each colored block represents a different component. Block size indicates the size of the component, and the shaded *waves* joining the blocks represent migrations. The width of the wave is proportional to the fraction of users that had migrated.

**Temporal Dynamics.** Here, we study the stability of the ASKEWBOWTIE by constructing time-slices of the conversation graph, $G(\mathcal{T}_{(d,d+1)})$, in cumulative fashion at different days during the promotional event. We see in

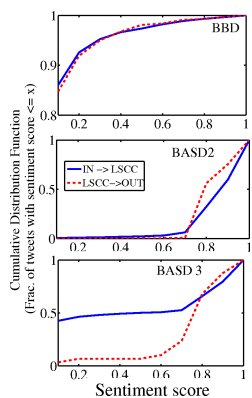

**Figure 3:** Cumulative Distribution Function of the sentiment-scores for two flows - $\mathcal{T}(\texttt{IN} \to \texttt{LSCC})$ and $\mathcal{T}(\texttt{LSCC} \to \texttt{OUT})$ in the ASKEWBOWTIE (in x-axis, $0$ and $1$ indicate extreme negative and extreme positive sentiments respectively.

Figure 2 that the observed skewness of the ASKEWBOWTIE structure remains invariant across the days. TENDRILS tend to be the most unstable among the components, with $35\%$ of the users inside it eventually moving to the rest of the components, majority towards IN. We notice that $51.8\%$ of the users who freshly join the conversations each day add to IN, followed by $37.2\%$ added to TENDRILS each day. In contrast, LSCC sees only $4.04\%$ of the arrivals each day, hinting at the relative stability of the core of the conversation. From the perspective of the organizers of these promotional events, the migration of users to LSCC and OUT indicates increasing engagement levels.

*Flows*
Now we turn our attention from the components to the flow across the components of the ASKEWBOWTIE. In particular, we seek flow features that discriminate the $\mathcal{T}(\texttt{IN} \to \texttt{LSCC})$ flow from the $\mathcal{T}(\texttt{LSCC} \to \texttt{OUT})$.

**Opinion.** In Figure 3, we plot the cumulative distribution function of the sentiment-scores of tweets $t \in \mathcal{T}(\texttt{IN} \to \texttt{LSCC})$ and $t \in \mathcal{T}(\texttt{LSCC} \to \texttt{OUT})$. We observe that the $\mathcal{T}(\texttt{IN} \to \texttt{LSCC})$ flow is more negative in general, barring BASD2 where the overall sentiment was positive, and BBD where the overall sentiment was predominantly negative. This conforms to our hypothesis that the UNHAPPY users that abound in IN are responsible for the more negative opinions in $\mathcal{T}(\texttt{IN} \to \texttt{LSCC})$.

**Formality.** We turn to study the *language* of the tweets - in particular, their degree of adherence to the linguistic standards. Using the technique proposed in [4], we answer the question: Is $\mathcal{T}(\texttt{LSCC} \to \texttt{OUT})$ more formal than $\mathcal{T}(\texttt{IN} \to \texttt{LSCC})$? Table 5 summarizes our observations and demonstrates that communications in $\mathcal{T}(\texttt{LSCC} \to \texttt{OUT})$ are *more* formal than the communications in $\mathcal{T}(\texttt{IN} \to \texttt{LSCC})$, again confirming our hypothesis. This perhaps indicates that LSCC serves as a refiner of the tweets flowing from IN to OUT and provides more meaningful information to the users in OUT.

**Conclusions and Future Work**
In this paper, we weaved together all the conversations around large-scale e-commerce promotional events in Twitter into a conversation graph, and found that the graph demonstrates a stable ASKEWBOWTIE structure. We believe that the insights regarding structure and temporal dynamics of the Conversation Graph obtained through this study would help design better Social Relationship Management tools, which is our penultimate goal.

**References**

[1] Andrei Broder, Ravi Kumar, Farzin Maghoul, Prabhakar Raghavan, Sridhar Rajagopalan, Raymie Stata, Andrew Tomkins, and Janet Wiener. 2000. Graph Structure in the Web. *Comput. Netw.* 33, 6 (2000), 309–320.
[2] Flavio Chierichetti, Jon Kleinberg, Ravi Kumar, Mohammad Mahdian, and Sandeep Pandey. 2014. Event detection via communication pattern analysis. In *AAAI*. 51–60.
[3] Peter Cogan, Matthew Andrews, Milan Bradonjic, W Sean Kennedy, Alessandra Sala, and Gabriel Tucci. 2012. Reconstruction and analysis of twitter conversation graphs. In *HotSocial*. ACM, 25–31.
[4] Yuheng Hu, Kartik Talamadupula, and Subbarao Kambhampati. 2013. Dude, srsly?: The Surprisingly Formal Nature of Twitter's Language. In *ICWSM*. Cambridge, USA, 244–253.
[5] Saif M Mohammad, Svetlana Kiritchenko, and Xiaodan Zhu. 2013. NRC-Canada: Building the state-of-the-art in sentiment analysis of tweets. In *SEM*. 321–327.